\documentstyle[twocolumn,pra,aps]{revtex}
          \begin{document}
          \draft
          \title{The Representation of Natural Numbers in Quantum
Mechanics}
          \author{Paul Benioff\\
           Physics Division, Argonne National Laboratory \\
           Argonne, IL 60439 \\
           e-mail: pbenioff@anl.gov}
           \date{\today}

          \maketitle
          \begin{abstract}
This paper represents one approach to making explicit some of the
assumptions and conditions implied in the widespread
representation of numbers by composite quantum systems. Any
nonempty set and associated operations is a set of natural
numbers or a model of arithmetic if the set and operations
satisfy the axioms of number theory or arithmetic. This work is
limited to $k-ary$ representations of length $L$ and to the
axioms for arithmetic modulo $k^{L}$. A model of the axioms is
described based on an abstract $L$ fold tensor product Hilbert
space ${\cal H}^{arith}$. Unitary maps of this space onto a
physical parameter based product space ${\cal H}^{phy}$ are then
described. Each of these maps makes states in ${\cal H}^{phy}$,
and the induced operators, a model of the axioms. Consequences of
the existence of many of these maps are discussed along with the
dependence of Grover's and Shor's Algorithms on these maps. The
importance of the main physical requirement, that the basic
arithmetic operations are efficiently implementable, is
discussed. This condition states that there exist physically
realizable Hamiltonians that can implement the basic arithmetic
operations and that the space-time and thermodynamic resources
required are polynomial in $L$.
\end{abstract}
           \pacs{03.65.-w,89.70+c,03.65.Bz}

\section{Introduction}
As is well known numbers play an essential role in physics and in
many other disciplines. The results of both experimental work and
theoretical computations are often given as numbers.  Comparison
of these numbers is essential to the validation process for any
physical theory such as quantum mechanics. As inputs to or
outputs of computations or experiments, numbers correspond to
states of physical systems.  From an information theoretic
viewpoint, this correspondence is essential as these states carry
information. As Landauer has emphasized, "Information is
Physical" \cite{Landauer}. This is taken very seriously here.

However, the fact that many states of many physical systems
correspond to numbers, has, for the most part, been assumed and
used implicitly. There has been little attempt to make explicit
the assumptions and conditions involved in representing numbers
by states of physical systems.

This paper represents one approach to making  some of the
assumptions and conditions explicit. The emphasis is on the
mathematical and physical aspects in the representation of
numbers by states of physical systems. No new models of
computation are presented. However making the assumptions
explicit does offer some insight into the importance of various
conditions that may not have been realized so far.  An example
(Section \ref{EIAO}) is the essential role played by the
condition that there exist physically realizable dynamical
operators that can efficiently implement basic arithmetic
operations. The fact that these conditions are satisfied for a
wide variety of systems, as shown by the ubiquitous existence of
computers, does not detract from their importance.

Such a study is also relevant to the development of a coherent
theory of mathematics and physics together, which, in one form or
another, is a goal of many physicists \cite{Tegm,Wein,BenDTVQM}.
Any such coherent theory must take account in detail of how
numbers are represented by states of physical systems.

In this paper considerations will be limited to quantum systems.
This is not a serious limitation because of the assumed universal
applicability of quantum mechanics or a related theory such as
quantum field theory.  In this case all physical systems are
quantum systems and all states of these systems are (pure or
mixed) quantum states.  This is the case whether the systems are
microscopic or macroscopic or whether macroscopic systems can be
described by classical mechanics.

For quantum systems numbers are represented by tensor products of
states of different degrees of freedom of a system. Usually the
system is composite with each degree of freedom associated with a
component system. For microscopic systems one condition systems
must satisfy is that they have states for which  the switching
time, $t_{sw}$, is short compared to the decoherence time
$t_{dec}$ or $t_{sw}\ll t_{dec}$ \cite{DiVincenzo}.  This is a
dynamic condition as it is based on the Hamiltonian for the
systems including their interaction with other systems and the
environment.

This condition eliminates many state spaces of microscopic quantum
systems for representation of numbers. A 2-dimensional example
would be the state space based on two highly excited states of
nuclei that have halflives short compared to $t_{sw}$. On the
other hand spin projection states of spin $1/2$ ground state
nuclei in in molecules in a magnetic field are suitable and are
used  in NMR quantum computers \cite{Gershenfeld,Favel,Chuang}.

Macroscopic quantum systems are such that $t_{sw}\gg t_{dec}$ for
all states of interest.  In this case the systems are candidates
for number representation for classical computation if the
systems have states that are stabilized by environmental
interactions for times long compared to the switching time.  The
widespread existence of macroscopic computers shows that both
$t_{sw}\gg t_{dec}$ and environmental stabilization occurs for
many quantum systems.

Because of the recent widespread interest in quantum computing,
the emphasis of this paper is on number representation by states
of microscopic quantum systems.  However most of the material
also applies to macroscopic systems.

The first step in giving an exact meaning to the representation
of numbers by tensor product states of quantum systems is to
specify exactly what natural numbers are.  Without such a
specification all computations are meaningless physical
operations. Here the axiomatic approach is used by defining a
nonempty set as a set of all natural numbers if it is a model of
the axioms of number theory or arithmetic
\cite{Shoenfield,Smullyan}. These axioms are discussed in the
next section along with changes needed to account for the
limitation of this paper to tensor product states with an
arbitrary but fixed finite number $L$ of components, or $k-ary$
representations of length $L$. The corresponding arithmetic
becomes arithmetic modulo $k^{L}$.

It is possible to model the axioms directly on a physical Hilbert
space ${\cal H}^{phy}$ describing a composite quantum system with
$L$ components.  However the literature on quantum computing
makes much use of product qubit states of the form
$|\underline{s}\rangle$ where $ \underline{s}$ is any function
from ${1,2,\cdots ,L}$ to $\{0,1\}$.  Since the Hilbert space of
these states  is a very useful reference base for discussing
quantum computation that is independent of any physical model,
this approach will be used here.

To this end a purely mathematical model of these axioms is
described in Section \ref{AHSM} that is based on a tensor product
Hilbert space ${\cal H}^{arith}=\otimes_{j=1}^{L}{\cal H}_{j}$ of
$L$ $k$ dimensional Hilbert spaces ${\cal H}_{j}$. Unitary
operators on this space are defined to correspond to the basic
arithmetic operations, successor, plus, and times, whose
properties are given by the axioms. The presence of $L$ successor
operators, one for each power of $k$, rather than just one as
described by the axioms, is based on the condition  of efficient
implementation discussed later on.

Tensor product states of physical properties of microscopic
composite quantum systems belonging to the Hilbert space ${\cal
H}^{phy}$ are discussed in  Section \ref{PHSM}. A priori these
states, as products over a label set $A$ of $L$ physical
parameter values, do not correspond to any number. Also operators
on these states are meaningless regarding any numerical
interpretation.

This is remedied by describing tensor product preserving unitary
operators from ${\cal H}^{arith}$ to ${\cal H}^{phy}$. For each
of these operators, ${\cal H}^{phy}$, along with induced
representations of the operators for the basic arithmetic
operations, becomes a model for the axioms of modular arithmetic.

So far nothing has been said about the physical realizability of
any of these models of the axioms. This is especially relevant
for the operators as they are many system nonlocal operators.
This is remedied in Section \ref{EIAO} where the important
condition of efficient implementability of the basic arithmetic
operations is described. In essence the condition requires that a
composite quantum system be such that there exist physically
realizable Hamiltonians that can implement the basic arithmetic
operations.  In addition the space-time and thermodynamic
resources required for implementation must be polynomial in $L$.
The importance of this condition rests in the fact that it is
additional to and independent of the axioms of arithmetic.  To see
this one notes that there are many models of the axioms that do
not satisfy this requirement.  A simple physical model is any one
based on an unary representation of the numbers as most
arithmetic operations are inefficient in this representation.

The question arises if use of  ${\cal H}^{arith}$ can be bypassed
by modeling the axioms directly on ${\cal H}^{phy}$ where ${\cal
H}^{phy}$ has an arbitrary tensor product structure. In general
this is possible as any structure satisfying the axioms is
acceptable. The discussion of this in Section \ref{IMHN} is based
on a description of properties of a set of operators indexed by a
set of physical parameters. The properties are also defined to
address the question of necessary and sufficient conditions to
conclude that ${\cal H}^{phy}$ must have a tensor product
structure suitable for length $L$ $k-ary$ representations of
numbers.

A final section discusses some other aspects and open questions
resulting from this work. The importance of the efficient
implementability condition in excluding most models of modular
arithmetic on ${\cal H}^{phy}$ is noted as are some aspects of
the use of numbers to describe $k-ary$ representations of length
$L$.

It must be emphasized that the work of this paper is one attempt
to make explicit the assumptions and conditions that are assumed
implicitly in the representation of numbers by states of quantum
systems and in work in the the literature on quantum computing.
Examples of this work are given in papers by Beckman et. al.
\cite{Beckman} and Vedral et.al.\cite{Vedral} that describe
networks of quantum gates to carry out basic arithmetic
operations. The description is in terms of unitary operators on
${\cal H}^{arith}$ (extended to include ancillary qubits) as
ordered products of polynomially many elementary gate operators.
The distinction between physical models, with the associated requirement of
efficient physical implementation, and mathematical models is not
maintained (and is not needed) in the papers. Also efficient
physical implementability implies more than minimizing the number
of ancillary qubits and restriction to polynomially many gate
operations.  These aspects are discussed more in Sections
\ref{AHSM} and \ref{EIAO}.

\section{The Axiomatic Description of Numbers}
The first step in making explicit what is involved in the
representation of numbers by quantum states is to define the
natural numbers. One method of doing this is to follow
mathematical logic and define any nonempty set to be a set of
natural numbers  if it is a model for the axioms of arithmetic or
number theory \cite{Shoenfield,Smullyan}.  A model for any axiom
system is a collection of elements in which all the axioms are
true.

Here the main interest is models of arithmetic based on Hilbert
spaces that are tensor products of an arbitrary but fixed number
of component spaces. As a result the axioms to be satisfied are
those for arithmetic modulo $N$ where $N$ is arbitrary but fixed.
This arithmetic satisfies some of the axioms for all natural
numbers. Others need to be either deleted or modified. It also
satisfies axioms for a commutative ring with identity
\cite{Adamson}.

The exact form and content of  axioms for modular arithmetic is
not important here. What is important is that both the arithmetic
and ring  axioms have in common the required existence of binary
operations $+$ and $\times$ with certain properties. Also an
unary successor operation  $S$ is required by the arithmetic
axioms.  The properties the binary operations must have include
commutativity, associativity, the existence of identities  $0$
and $S(0)$ for $+$ and $\times$, and the distributivity of
$\times$ relative to $+$. Also $S$ commutes with $+$ and $x\times
S(y)=x\times y+x$ for all $x,y$.\footnote{The importance of these
axioms lies in the requirement of the existence of binary
operations $+,\times$ with certain properties. The fact that some
of the  axioms may be redundant is of no importance here.} The
arithmetic axioms defining an order relation and the induction
schema are not considered  as they are not needed for the
purposes of this paper. However it is useful to keep in mind that
the ordering axioms establish the discreteness of the natural
numbers in the sense that there is no number between $x$ and its
successor $S(x)$.

\section{Abstract Hilbert Space Models} \label{AHSM}
The next step is the description of a purely mathematical model
of these axioms based on whatever mathematical systems are
appropriate for the physical systems being considered.  Since
interest here is in $k-ary$ representations of length $L$ of
natural numbers for composite quantum systems, a model based on
an abstract Hilbert space ${\cal H}^{arith}$ is needed. To this
end let ${\cal H}^{arith} = \otimes_{j=1}^{L}{\cal H}_{j}$ be  an
$L$ fold tensor product Hilbert space where ${\cal H}_{j}$ is a
$k$ dimensional Hilbert space.

For each $j$, the basis states of interest in ${\cal H}_{j}$ have
the form $|\ell ,j\rangle$ where $j$ denotes the label or
property characterizing a qubyte  and $\ell =0,1,\cdots ,k-1$. A
product state basis in $\cal H$ can be given in the form
$|\underline{s}\rangle =
\otimes_{j=1}^{L}|\underline{s}(j),j\rangle$ where
$\underline{s}$ is any function from $1,\cdots ,L$ to $0,\cdots
,k-1$. (Here qubits or qubytes \cite{Schumacher,Fazio} refer to
quantum bits or bytes of information for $k=2$ or $k\geq 2$
respectively.)

The presence of the parameter $j$ in the state and not as a
subscript, as in $\otimes_{j=1}^{L}|\underline{s}(j)\rangle_{j}$,
is required as the action of  operators corresponding to the
basic arithmetic operations depends on the value of $j$.  It is
not possible to express this dependence if $j$ appears as a
subscript of $\rangle$ and not between $|$ and $\rangle$.

An important function of the axioms is to provide properties of
the unary operation $S$ and the binary operations $+$ and
$\times$.   For reasons based on efficient implementation
(Section \ref{EIAO}), it is quite useful to define $L$ different
successor operators, $V^{+1}_{j}$, for $j=1,\cdots ,L$. These
operators are defined to correspond to the addition of $k^{j-1}
\bmod k^{L}$ where $V^{+1}_{1}$ corresponds to $S$ in the axioms.
These operators and those for $+$ and $\times$ correspond to the
basic arithmetic operations.

It is to be emphasized that definitions of the $+,$ and $\times$
are given to show their dependence on the $V^{+1}_{j}$. Also they
are required by the axioms of arithmetic. The purpose is
definitely not to present the definitions as something new as
these operators are widely used.

For instance the widely discussed networks of quantum gates are
examples of the abstract models considered here for $k=2$.  In the
networks the  states in ${\cal H}^{arith}$ are represented by
horizontal qubit lines and ordered products of gate operators
represent operators in $B({\cal H}^{arith})$.  Specific examples
of this for the basic arithmetic operations of addition,
multiplication and modular exponentiation are described in
\cite{Beckman,Vedral}.

It is also the case that in many physical models space and time
directions can be assigned to the abstract networks.  In this
case the spatial ordering of the qubit lines is part of any
mapping of the abstract models based on ${\cal H}^{arith}$ to
physical models based on ${\cal H}^{phy}$ in which the
corresponding component systems are distinguished by spatial
positions. These mappings are examples of the  mappings "$g$"
discussed in Section \ref{PHSM}. The time ordering of the quantum
gates corresponds to mapping the ordering of gate operators in the
abstract model to a time ordered product of physically
implementable quantum gate operators. This is part of the
requirement of physical implementability (Section \ref{EIAO}).

\subsection{Definitions of the $V^{+1}_{j}$} \label{DV}
The definition of the $V^{+1}_{j}$ is straightforward.  For each
$j$ let $u_{j}$ be a cyclic shift \cite{Halmos} of period $k$ that
acts on the states $|\ell,j\rangle$ according to
$u_{j}|\ell,j\rangle=|\ell +1\bmod k,j\rangle$. $u_{j}$ is the
identity on all states $|m,j^{\prime}\rangle$ where
$j^{\prime}\neq j$.  Define $V^{+1}_{j}$ by
\begin{equation}
V^{+1}_{j} = \left\{ \begin{array}{ll} u_{j}P_{\neq (k-1),j} +
V^{+1}_{j+1}u_{j}P_{(k-1),j} & \mbox{if $1\leq j <
L$} \\
u_{L} & \mbox{if $j=L$} \end{array} \right. \label{p1impl}
\end{equation}
Here $P_{(k-1),j} =|k-1,j\rangle\langle k-1,j| \otimes 1_{\neq
j}$ is the projection operator for finding  the $j$ component
state $|k-1,j\rangle$ and the other components in any state.
$P_{m,j}$ and $u_{j}$ satisfy the commutation relation
$u_{j}P_{m,j}=P_{m+1,j}u_{j} \bmod k$ for $m=0,\cdots ,k-1$. Also
$P_{(\neq k-1),j} = 1-P_{(k-1),j}$. This follows from the fact
that the label spaces for each qubyte are one dimensional so that
the operator $1_{\neq j}\otimes |j\rangle\langle j|\otimes 1_{\neq
j}$ is the identity on the Hilbert space spanned by the $k^{L}$
states $|\underline{s}\rangle$

This definition is implicit in that $V^{+1}_{j}$ is defined in
terms of $V^{+1}_{j+1}$.  An explicit definition is given by
\begin{eqnarray}
V^{+1}_{j} & = & \sum_{n=j}^{L}u_{n}P_{(\neq k-1),n} \prod_{\ell
=j}^{n-1}u_{\ell}P_{(k-1),\ell}
\nonumber \\
& & \mbox{} + \prod_{\ell = j}^{L} u_{\ell}P_{(k-1),\ell}
\label{p1expl}
\end{eqnarray}
In this equation the unordered product is used because for any
$p,q$, $u_{m}P_{p,m}$ commutes with $u_{n}P_{q,n}$ for $m\neq n$.
Also for $n=j$ the product factor with $j\leq \ell \leq n-1$
equals $1$.

There are two basic properties the operators $V^{+1}_{j}$ must
have: they  are cyclic shifts and, for each $j<L$, they satisfy
 \begin{equation}
(V^{+1}_{j})^{k} = V^{+1}_{j+1} \label{logeff}
\end{equation}
Also if $j=L$ then $(V^{+1}_{L})^{k} = 1$. To show that
$V^{+1}_{j}$ is a shift, let $|\underline{s} \rangle$ be a
product  state such that for each $m=1,2,\cdots ,L$ the
component  states $|\underline{s}_{m}\rangle,
u_{m}|\underline{s}\rangle,(u_{m})^{2}|\underline{s}\rangle,
\cdots , (u_{m})^{k-1}|\underline{s}\rangle$ are pairwise
orthonormal. It then follows from Eq. \ref{p1expl} and the
properties of the $u_{m}$ that any product state
$|\underline{s}\rangle$ is orthogonal to the state
$V^{+1}_{j}|\underline{s}\rangle$ and that $V^{+1}_{j}$ is norm
preserving on these states.

Assume that Eq. \ref{logeff} is valid. Then for each $j$
$(V^{+1}_{j})^{k^{L-j+1}} = 1$. This, and the facts that for all
tensor product  states $|\underline{s}\rangle ,\;
V^{+1}_{j}|\underline{s}\rangle$ is also a tensor product state
which is orthogonal to $|\underline{s}\rangle$,  show that
$V^{+1}_{j}$ is a cyclic shift.  The existence of a tensor
product basis that is common to all the $V^{+1}_{j}$ follows from
Eq. \ref{logeff}.

To prove Eq. \ref{logeff} it is easiest to use Eq. \ref{p1impl}.
Since  $V^{+1}_{j+1}$ commutes with $u_{\ell}P_{n,\ell}$ for all
$\ell \leq j$ and the commutation relations $P_{\neq n,j}u_{j} =
u_{j}P_{\neq (n-1),j}$ and $P_{n,j}u_{j} = u_{j}P_{(n-1),j}$
hold, one has for each $m\leq k$
\begin{displaymath}
(V^{+1}_{j})^{m} = (u_{j})^{m}\prod_{\ell =1}^{m}P_{\neq
(k-\ell),j} + V^{+1}_{j+1}(u_{j})^{m}(\sum_{\ell = 1}^{m}
P_{(k-\ell),j}).
\end{displaymath}
Here $P_{\neq n,j} = 1 - P_{n,j}$.  For $m=k$ the term with the
product of the projection operators gives $0$ and the sum of the
projection operators gives unity. The desired result follows from
the fact that $(u_{j})^{k} =1$. Also $(V^{+1}_{L})^{k} = 1$
follows directly from the definition of $V^{+1}_{L}$.

The above shows that informally the action of $V^{+1}_{j}$
corresponds to addition $\bmod k^{L}$ of $k^{j-1}$ on the product
basis. This cannot yet be proved as addition $\bmod k^{L}$ has
not yet been defined. Also the adjoint $(V^{+1}_{j})^{\dagger}$
of $V^{+1}_{j}$ corresponds informally to subtraction $\bmod
k^{L}$ of $k^{j-1}$.  This can be seen from the fact that
$(V^{+1}_{j})^{\dagger}V^{+1}_{j} = 1$ where
\begin{eqnarray}
(V^{+1}_{j})^{\dagger} & = & \sum_{n=j}^{L}P_{(\neq
k-1),n}u^{\dagger}_{n} \prod_{\ell =j}^{n-1}
P_{(k-1),\ell}u^{\dagger}_{\ell} \nonumber \\
& & \mbox{} + \prod_{\ell =
j}^{L}P_{(k-1),\ell}u^{\dagger}_{\ell}. \label{m1expl}
\end{eqnarray}
This result is obtained using the commutativity of the shifts and
projection operators for different component systems.

It should be noted that the operators $V^{+1}_{j}$ play an
important role in quantum computation.  This is the case even
though for each product state $|\underline{s}\rangle$ the state
$V^{+1}_{j}|\underline{s}\rangle$ is also a product state and is
not a linear superposition of these states.  The importance comes
from the fact that these operators along with their efficient
implementation are used to define the basic arithmetic operations
for a quantum computer and to carry out quantum algorithms.  For
example in Shor's factoring quantum algorithm \cite{Shor}, they
are used in the step in which the function $f_{y}({s}) =
y^{s}\bmod N$ is calculated for each component state
$|\underline{s}\rangle$.

\subsection{Plus}
\label{Plus} It is straightforward to define the plus ($+$)
operation in terms of the $V^{+1}_{j}$. To ensure unitarity the
definition will be based on states of the form
$|\underline{s,w}\rangle =|\underline{s}\rangle \otimes
|\underline{w}\rangle$ that describe two $L$ qubyte product
states.

To define the $+$ operation let $V^{+\ell}_{j}=
(V^{+1}_{j})^{\ell}$ represent $\ell$ iterations of $V^{+1}_{j}$.
Then $+$ is defined by
\begin{eqnarray}
+ |\underline{s}\rangle \otimes |\underline{w}\rangle & = &
|\underline{s}\rangle \otimes V^{+s_{L}}_{L} V^{+s_{L-1}}_{L-1}
\cdots V^{+s_{2}}_{2}V^{+s_{1}}_{1} |\underline{w}\rangle
\nonumber \\
& & \mbox{} = |\underline{s},\underline{s+w}\rangle \label{plus}
\end{eqnarray}
Here the numeral expression $|\underline{s+w}\rangle$ is defined
to be that generated from $\underline{w}\rangle$ by the action of
the product $\prod_{j=1}^{L}V^{+s_{j}}_{j}$. Note that the
different $V^{+1}_{j}$ commute.

For pairs of product states, which are first used here, the
domains of the functions $ \underline{s}$ and $ \underline{w}$
must be different.  This is based on the requirement that an
algorithm must be able to distinguish components of
$|\underline{s}\rangle$ from components of
$|\underline{w}\rangle$.  This can be achieved by setting
$|\underline{s}\rangle \otimes |\underline{w}\rangle
=|\underline{s\ast w}\rangle$ where $\underline{s\ast w}$ denotes
the concatenation of $ \underline{w}$ to $ \underline{s}$.  That
is, $\underline{s\ast w}$ is a function from $1,\cdots ,2L$ to
$0,\cdots ,k-1$ where $\underline{s\ast w}(h) = \underline{s}(h)$
for $h\leq L$ and $\underline{s\ast w}(h) =\underline{w}(h-L)$
for $h> L$.

As defined the $+$ operator is unitary on the Hilbert space
spanned by all pairs of length $L$ numeral expression states.
Thus a reversible implementation of it is possible where the
procedure makes use of the procedures  for implementing the
$V^{+1}_{j}$. Eq. \ref{plus} shows that the procedure can be
carried out by carrying out, for each $j=1,2,\cdots ,L$, $s_{j}$
iterations of $V^{+1}_{j}$ where $s_{j}$ is the number $
\underline{s}(j)$ associated with the qubyte state
$|\underline{s}(j),j \rangle$ in
$|\underline{s}\rangle=\otimes_{j=1}^{L}|\underline{s}(j),j\rangle$.
Since $+$ is unitary, so is the adjoint, $+^{\dagger}$. Since $+$
was defined to correspond to addition modulo $k^{L}$, the adjoint
corresponds to subtraction modulo $k^{L}$. That is if
$+|\underline{s}\rangle \otimes |\underline{w}\rangle
=|\underline{s}\rangle \otimes |\underline{s+w}\rangle$ then
$+^{\dagger}|\underline{s}\rangle \otimes |\underline{s+w}\rangle
= |\underline{s}\rangle \otimes |\underline{w}\rangle$.

\subsection{Times}
\label{T} Here a definition of multiplication is given that is
based on efficient iteration of $+$ and is similar to the method
taught in primary school.  The method is efficient relative to
that for $+$.

Reversibility of the operations requires that the operator
$\times$ be unitary. (Caution: the adjoint of $\times$ is not
division.) This means that both input product  states and the
product state with the result must be preserved. It is also
convenient to have one extra product state for storing and acting
on intermediate results.  This state begins and ends as
$|\underline{0}\rangle$. For initial states of the form,
$|\underline{s},\underline{w},\underline{0}, \underline{0}\rangle
=|\underline{s}\rangle\otimes |\underline{w}\rangle\otimes
|\underline{0}\rangle\otimes |\underline{0}\rangle$,
\begin{equation}
\times |\underline{s},\underline{w},\underline{0} ,\underline{0}
\rangle = |\underline{s},\underline{w},
\underline{0},\underline{s\times w}\rangle \label{times}
\end{equation}
where $|\underline{s\times w}\rangle$ is the state resulting from
the action of $\times$. It is supposed to correspond to the
result of multiplying, $\bmod {k^{L}}$, the numbers corresponding
to the states $|\underline{s}\rangle$ and $\underline{w}\rangle$.

In order to define $\times$ explicitly one needs to be able to
generate the states $|\underline{k^{j-1}\times w}\rangle$
corresponding to multiplication of $w$ by $k^{j-1}$. For each
$j=1,\cdots ,L$ these states are added to themselves $s_{j}$
times.  The final result is obtained by adding all the resulting
states so obtained. Details are provided in the Appendix.

\subsection{Required Properties of the $V^{+1}_{J}$, Plus, Times}
\label{reqprop}  As was noted the operators $V^{+1}_{j},+,\times$
must satisfy the properties expressed by the axioms for modular
arithmetic. These include the axioms for arithmetic
\cite{Shoenfield,Smullyan} modified for modularity and the
p[resence of $L$ successors, and possibly axioms for a commutative
ring with identity \cite{Adamson}.

Properties that must be satisfied include that expressed by Eq.
\ref{logeff} and the requirements that the successor operations
commute with $+$, (i.e. $+(1\otimes V^{+1}_{j}) = (1\otimes
V^{+1}_{j})+$), the existence of additive and multiplicative
identities, which are the states $|\underline{0}\rangle$ and
$|\underline{1}\rangle = V^{+1}_{1}|\underline{0}\rangle$, and
the distributivity of $\times$ over $+$. Also $+$ and $\times$
are associative and commutative.

Proof of these properties from the definitions and Eq.
\ref{logeff}, which has already been proved, is straight forward
and will not be given here. Note that the proofs of some of the
properties do use the corresponding properties  of the numbers
appearing in the exponents.  For example to prove that addition
is commutative, $|\underline{s+w}\rangle = |\underline{w+s}
\rangle$, Eqs. \ref{plus} and \ref{p1expl} give
$|\underline{s+w}\rangle = \prod_{h=1}^{L}
(V^{+1}_{j})^{s_{h}+w_{h}}|\underline{0}\rangle$ and
$|\underline{w+s}\rangle = \prod_{h=1}^{L}
(V^{+1}_{j})^{w_{h}+s_{h}}|\underline{0}\rangle$. The equality of
these two states follows from $s_{h}+w_{h} = w_{h}+s_{h}$ for
each $h$.

\section{Physical Hilbert Space Models} \label{PHSM}
The Hilbert space models described so far are purely abstract in
that they do not refer to any physical properties.  They do
however, serve as a common reference point for models based on
physical properties of physical systems. They also give a useful
method to associate numbers with quantum states of these systems.

To begin, let  $A$ and $B$  be sets of $L$ and $k$  different
physical parameters or values of some physical properties or
observables $\hat{A}$ and $\hat{B}$. The $A$ parameters are used
to distinguish or label different components of a composite
quantum system and $B$ is a set of values of a different physical
property associated with each component system.  For example $A$
could be a set of $L$ arbitrary locations of component spin $1/2$
systems on a 2 dimensional surface and $B=\{\uparrow
,\downarrow\}$ denoting spin  aligned along or opposite some axis
of quantization. Another example, representative of NMR quantum
computation \cite{Gershenfeld,Favel,Chuang}, has $A$ as a set of
hyperfine splittings of nuclear spin states and $B= \{\uparrow
,\downarrow \}$. Here the values of $A$ must contain sufficient
information so the physical process can distinguish between the
different nuclear spins.

Let $ \underline{t}$ be any function from $A$ to $B$ and
$|\underline{t}\rangle= \otimes_{a\epsilon A}
|\underline{t}(a),a\rangle$ be the corresponding tensor product
state. Let ${\cal H}^{phy} =\otimes_{a\epsilon A} {\cal H}_{a}$
be the $k^{L}$ dimensional Hilbert space spanned by all the
states $|\underline{t}\rangle$.  Each ${\cal H}_{a}$ is a $k$
dimensional Hilbert space spanned by states of the form
$|h,a\rangle$ where $h\epsilon B$.

The presence of $a$ as a separate part in each component state
$|\underline{t}(a), a\rangle$, and not as a state subscript as in
$|\underline{t}(a)\rangle_{a}$, is essential as an algorithm uses
the value of $a$ to distinguish the different component systems.
This is based on the view that the state of the composite quantum
system contains all the quantum information available to the
algorithm. In particular the states must contain sufficient
information so that the algorithm can distinguish among the
component systems.  This is especially the case for any algorithm
whose dynamics is described by a Hamiltonian that is selfadjoint
and time independent. This is an example of Landauer's dictum
"Information is Physical" \cite{Landauer}.

This description can be generalized in that the physical property
observable $\hat{B}$ of the component systems can depend on the
values of $a$ in $A$. An example this, which also has different
component systems  replaced by different degrees of freedom of
one system, is shown by an ion trap example \cite{Monroe}.  Here
the states of one  degree of freedom are the ground and first
excited state of the ion in the harmonic well trap. The
corresponding states of the other are the ground and first
excited electronic state of the ion. This type of generalization
will not be pursued here.

\subsection{Representation of Numbers and Arithmetic Operations in
${\cal H}^{phy}$} The goal here is for states in ${\cal H}^{phy}$
to represent numbers. However, it is clear that, a priori,
neither the product states $|\underline{t}\rangle=
\otimes_{a\epsilon A} |\underline{t}(a),a\rangle$ nor linear
superpositions of these states  represent numbers.  For the
$|\underline{t}\rangle$ the reason is that there is no
association between the labels $a$ and powers of $k$; also there
is no association between the range set $B$ of $\underline{t}$
and the numbers $0,1,\cdots ,k-1$.

This can be remedied by use of unitary maps from ${\cal
H}^{arith}$ to ${\cal H}^{phy}$ that preserve the tensor product
structure.  One way of doing this is to let $g$ and $d$ be any
bijections (one-one onto) maps from $1,2,\cdots ,L$ to $A$ and
from $0,1,\cdots ,k-1$ to $B$.  For each pair $g,d$ and each $j$
there is a corresponding unitary operator $w_{g,d,j}$ that maps
states $|h,j\rangle$ in ${\cal H}_{j}$ where $0\geq h \geq k-1$
to states in ${\cal H}_{g(j)}$ according to $w_{g,d,j}|h,j\rangle
= |d(h),g(j)\rangle$. This induces a unitary operator
$W_{g,d}=\otimes_{j=1}^{L}w_{g,d,j}$ from the product space
${\cal H}^{arith}$ to ${\cal H}^{phy}$ where
\begin{eqnarray}
W_{g,d}|\underline{s}\rangle = & \otimes_{j=1}^{L}w_{g,d,j}
|\underline{s}(j),j\rangle \nonumber \\ \mbox{} = &
\otimes_{j=1}^{L} |d(\underline{s}(j)),g(j)\rangle
=|\underline{s}_{g}^{d}\rangle. \label{wgddef}
\end{eqnarray}
Here $|\underline{s}_{g}^{d}\rangle$ is the physical parameter
based state in ${\cal H}^{phy}$ that corresponds, under $W_{g,d}$
to the number state $|\underline{s}\rangle$ in ${\cal H}^{arith}$.

This process can be inverted, using the adjoint
$W^{\dagger}_{g,d}$ to relate physical parameter states in ${\cal
H}^{phy}$ to number states in ${\cal H}^{arith}$. One has
\begin{eqnarray}
W^{\dagger}_{g,d}|\underline{t}\rangle = & \otimes_{a\epsilon
A}w^{\dagger}_{g,d,g^{-1}(a)} |\underline{t}(a),a\rangle \nonumber \\
\mbox{} = & \otimes_{a\epsilon A}
|d^{-1}(\underline{t}(a)),g^{-1}(a)\rangle
=|\underline{t}_{g^{-1}}^{d^{-1}}\rangle. \label{wdaggddef}
\end{eqnarray}
Here $|\underline{t}_{g^{-1}}^{d^{-1}}\rangle$ is the number state
in ${\cal H}^{arith}$ corresponding to the physical state
$|\underline{t}\rangle$. Note that
$W^{\dagger}_{g,d}=W_{g^{-1},d^{-1}}$ where $g^{-1},d^{-1}$ are
the inverses of $g$ and $d$, and $w_{g^{-1},d^{-1},a} =
w^{\dagger}_{g,d,g^{-1}(a)}$.

The operators $W_{g,d}$ also induce representations of the
$V^{+1}_{j},\; +,$ and $\times$ operators on the physical
parameter states in ${\cal H}^{phy}$. For the $V^{+1}_{j}$ one
defines $V^{d,+1}_{g,j}$ by
\begin{equation}
V^{d,+1}_{g,j} = W_{g,d}V^{+1}_{j}W^{\dagger}_{g,d}.
\end{equation}
An equivalent definition can be given by direct reference to the
maps $g,d$ and the operators $w_{g,d,j}$:
\begin{eqnarray}
V^{d,+1}_{g,j} & = & \sum_{n=j}^{L}u_{g(n)}^{d}P_{ \neq
d(k-1),g(n)} \prod_{\ell =j}^{n-1}u_{g(\ell )}^{d}P_{d(k-1),g(\ell
)}
\nonumber \\
&  & \mbox{} + \prod_{\ell = j}^{L} u_{g(\ell
)}^{d}P_{d(k-1),g(\ell )}. \label{p1fexpl}
\end{eqnarray}
Here $P_{d(k-1),g(\ell)} =
w_{g,d,\ell}P_{k-1,\ell}w^{\dagger}_{g,d,\ell}$ and $u_{g(\ell )}
= w_{g,d,\ell}u_{\ell }w^{\dagger}_{g,d,\ell}$.

In a similar fashion one can use the $W_{g,d}$ to define the
operator $+_{g,d}$ acting on the physical parameter states in
${\cal H}^{phy}\otimes {\cal H}^{phy}$.  The definition is based
on that given for the operator $+$ acting on ${\cal
H}^{phy}\otimes {\cal H}^{phy}$ Eq. \ref{plus}.  One has
\begin{equation}
+_{g,d}=(W_{g,d}\otimes W_{g,d})+(W^{\dagger}_{g,d}\otimes
W^{\dagger}_{g,d}). \end{equation} The operator $\times_{g,d}$ is
defined similarly from  $\times$ as defined in the Appendix.

It is clear from the above that there is no unique correspondence
between states in the arithmetic and physical Hilbert spaces.
There are $L!$ possible bijections $g$ and $k!$ possible
bijections $d$. Thus some or many of the $L!k!$ unitary operators
$W_{g,d}$ associate a different physical parameter state
$|\underline{s}^{d}_{g}\rangle$ with the number state
$|\underline{s}\rangle$. Conversely the $g$ and $d$ dependence of
$W^{\dagger}_{g,d}$ shows that many different number states
$|\underline{t}^{d}_{g}\rangle$ can be associated with the
physical state $|\underline{t}\rangle$. The multiplicity of these
correspondences depends on the states $|\underline{s}\rangle$ or
$|\underline{t}\rangle$ and the choices of $g$ and $d$.

It follows from the unitarity of $W_{g,d}$ that if the operators
$V^{+1}_{j},\; +,\; \times$ and the states
$|\underline{s}\rangle$ in ${\cal H}^{arith}$ satisfy the axioms
of modular arithmetic, then so do the operators
$V^{d,+1}_{g,j},\; +_{g.d},\; \times_{g,d}$ and states
$|\underline{s}^{d}_{g}\rangle$ in ${\cal H}^{phy}$. In this way
all the states $|\underline{s}^{d}_{g}\rangle$ in ${\cal
H}^{phy}$ and the operators
$V^{d,+1}_{g,j},\;+_{g,d},\;\times_{g.d}$ are a model of the
axioms of modular arithmetic.  The fact that superposition of the
states $|\underline{s}^{d}_{g}\rangle$ plays an important role in
quantum computation does not affect this conclusion.

This argument also applies to any unitary map $U$ from ${\cal
H}^{arith}$ to ${\cal H}^{phy}$ independent of whether $U$ is
tensor product preserving or not. However most of these maps are
not of interest because the operators $UV^{+1}_{j}U^{\dagger}$
are not physically implementable (Section \ref{EIAO}). Also the
states $U|\underline{s}\rangle$ nay not be stable or even
preparable.

\subsection{Grover's and Shor's Algorithms} \label{GSA}
Since the spaces ${\cal H}^{arith}$ and ${\cal H}^{phy}$, and
arithmetic models constructed on these spaces are unitarily
equivalent, one might think that dynamically an algorithm is
independent of the unitary map used.  This is not true in general
even if one restricts the maps to have the form of $W_{g,d}$: some
algorithms are independent of these maps and others are not.

To see this one notes that dynamically any quantum algorithm
carried out on a composite physical system must be sensitive to
the values of the physical parameters for the system.  This means
that the physical dynamics of an algorithm must be described by
some evolution operator acting on the states in ${\cal H}^{phy}$
or some other physical model of the system states. The physical
dynamics is not described on ${\cal H}^{arith}$.

It follows that any algorithm that can be described in terms of
states based on physical parameters is independent of the unitary
maps $W_{g,d}$. The dynamics does not depend on these maps
because what number a physical state represents is irrelevant to
the algorithm.  On the other hand, algorithms that compute
numerical functions must be described on ${\cal H}^{arith}$ as
number is of the essence for these.  It follows that the dynamics
of these algorithms depends on the maps $W_{g,d}$.

Grover's Algorithm \cite{Grover} and Shor's Algorithm \cite{Shor}
are examples of the two types of algorithm. Grover's Algorithm
corresponds to a quantum search of a set of data where each
element of the data base corresponds to a quantum state. The goal
is to find the one unknown but unique state with some property
different from the others. Here the quantum state representing
each data element will be taken to be a tensor product of qubit
states. This is not necessary, as Lloyd \cite{Lloyd} has shown.
However, the price for this is the need for an exponential
overhead of resources.

Here the relevant feature of Grover's Algorithm is that it can be
both defined and implemented on ${\cal H}^{phy}$ with no
reference to numbers represented by states in ${\cal H}^{arith}$.
To see this let $k=2$ and $B =\{\uparrow ,\downarrow\}$ for spin
up, spin down. The initial state can be written as $\psi
=(1/\sqrt{N})\sum_{ \underline{t}}|\underline{t}\rangle$ where
$|\underline{t}\rangle = \otimes_{a\epsilon A}
|\underline{t}(a),a\rangle$ and $N=2^{L}$.

Dynamically Grover's Algorithm \cite{Grover}  consists of
iterations of the unitary operator $-WI_{\underline{\uparrow}}W
I_{\underline{t_{u}}}$ on ${\cal H}^{phy}$. Here
$I_{\underline{\uparrow}}=1-2|\underline{\uparrow}
\rangle\langle\underline{\uparrow}|$ where
$|\underline{\uparrow}\rangle$ is the state with all $L$ systems
in the $|\uparrow \rangle$ state.  $I_{\underline{t_{u}}} =
1-2|\underline{t_{u}}\rangle\langle\underline{t_{u}}|$ and $W$ is
the Walsh Hadamard transformation. Here $|\underline{t_{u}}
\rangle$ is the unknown product state that is to be amplified,
and $W=\otimes_{a\epsilon A}(1/\sqrt{2})(\sigma_{x}+
\sigma_{z})_{a}$ is a tensor product of single qubit operators.
The $\sigma_{x},\; \sigma_{z}$ are the Pauli spin operators and
$\psi = W|\underline{\uparrow}\rangle$.

Shor's Algorithm \cite{Shor} for finding the two prime factors of
a large number is quite different in that it is essential that
the tensor product states represent numbers. This can be seen
from the steps of the algorithm
\begin{eqnarray}
\lefteqn{\frac{1}{\sqrt{N}}\sum_{\underline{s}}|\underline{s}\rangle
|\underline{i}\rangle  \Longrightarrow
\frac{1}{\sqrt{N}}\sum_{\underline{s}}|\underline{s}\rangle
|\underline{f_{m}(s)}\rangle} \nonumber \\
& \Longrightarrow & \frac{1}{N}
\sum_{\underline{w}}|\underline{w}\rangle \sum_{\underline{s}}
e^{-2\pi iws/N} |\underline{f_{m}(s)}\rangle
 \label{Shor1}
\end{eqnarray}
Here $|\underline{i}\rangle$ is the  initial product state,
usually shown as a constant sequence of $0s$. $f_{m}$ is a
numerical function defined by $f_{m}(x) =m^{x} \bmod M$ where m
and $M$ are relatively prime. The number $M$, which is to be
factored, and $N$ are related by $M^{2}\leq 2^{N}\leq 2M^{2}$
\cite{Shor,Miquel}.

Eq. \ref{Shor1} shows that the dynamics of Shor's algorithm can be
initially formulated as a unitary step operator $U_{Sh}$ acting on
${\cal H}^{arith}$. However, physically, the dynamics is
represented by the operator $W_{g,d}U_{Sh}W^{\dagger}_{g,d}$
acting on ${\cal H}^{phy}$. This shows that physically the
dynamical implementation of Shor's Algorithm depends on the
numberings $g$ and $d$ of the physical parameter sets $A$ and $B$.

More generally the requirement that the numerical function
calculated by the algorithm be invariant under any unitary map
from ${\cal H}^{arith}$ to ${\cal H}^{phy}$ means that the
physical implementation of the algorithm depends on the unitary
map.  For example, let $W_{g,d}$ be a unitary map as defined by
Eq. \ref{wgddef} and $A$ be a set of space locations of spin
$1/2$ systems with spin up $(\uparrow)$ spin down $(\downarrow)$
representing (through $d^{-1}$) $0,1$. Then the algorithm
dynamics clearly depends on $g$ as $g$ determines which space
location is associated with which power of $2$. A similar
argument holds for the dynamics dependence on $d$. Also the
correct interpretation of the measurement of the output depends
on both $g$ and $d$.

\section{Efficient Implementability of Arithmetic Operations}
\label{EIAO} Probably the most important requirement is that of
efficient implementablity of basic arithmetic operations. This
means that, for states of a physical system to represent numbers,
it must be possible to physically implement these operations and
the implementation must be efficient. This includes at least the
operations described by the axioms as efficient implementation of
these is a necessary condition for states of a quantum system to
represent numbers.

In the case of the $V^{d,+1}_{g,j}$ physical implementability
means there must exist a physically realizable Hamiltonian
$H_{g,j}^{d}$ such that for some time $t_{j}$,
$U_{g,j}^{d}(t_{j})=e^{-iH_{g,j}^{d}t_{j}}$ corresponds to
carrying out $V^{d,+1}_{g,j}$ on the states of the system. As
$V^{d,+1}_{g,j}$ is unitary, one has
$e^{-iH_{g,j}^{d}t_{j}}=V^{d,+1}_{g,j}$.  The presence of the
indices $d,g$ shows the dependence of $H^{d}_{g,j}$  on the
$W_{g,d}$.

Efficient implementation means that the time $t_{j}$ must be
short. For microscopic systems this is equivalent to the
condition that $t_{j}$ must be less than the decoherence time
$t_{dec}$.  If the Hamiltonian and system are such that
$V^{d,+1}_{g,j}$ is carried out in a number $n_{j}$ of basic
switching steps of duration $\Delta$, then  $n_{j} = t_{j}/\Delta
<t_{dec}/\Delta$ \cite{DiVincenzo} must hold.

For macroscopic systems the efficiency requirement is different
as $t_{dec}<<\Delta$. In this case $n_{j}$ must be polynomial and
not exponential in $L$. This means that $n_{j} = O(L^{c})$ with
$c\geq 0$ and $c$ not too large. $O()$ means "of the order of".

The efficiency requirement is much stricter for microscopic
systems than for macroscopic ones. The reason is that for most
systems $t_{dec}$ is small \cite{DiVincenzo}. This is one reason
why quantum computers are so hard to implement compared to
macroscopic computers. However, the requirement that $n_{j}$ be
polynomial in $L$ would also apply to any microscopic system for
which $t_{dec}/\Delta$ is very large, (e.g. $t_{dec}$ is several
hours or even longer).

The above is rather general in that it assumes that for each $j$
there is a distinct Hamiltonian $H_{g,j}^{d}$ to implement
$V^{d,+1}_{g,j}$. However for many systems all the
$V^{d,+1}_{g,j}$ may be implemented by just one Hamiltonian
$H_{g}^{d}$ with the different values of $j$ expressed by
different states of some ancillary systems.

The requirement of efficient implementation is the reason that
the $V^{d,+1}_{g,j}$ are defined separately for each $j$ rather
than defining them from $V^{d,+1}_{g,1}$ by $V^{d,+1}_{g,j} =
(V^{d,+1}_{g,1})^{k^{j-1}}$. Here $V^{d,+1}_{g,1}$ corresponds to
the successor operation $"+1"$ in axiomatic arithmetic
\cite{Shoenfield,Smullyan}. The exponential dependence on $j$
shown by this equation shows that if  efficient implementation
were required just for $V^{d,+1}_{g,1}$, then carrying out of the
$V^{d,+1}_{g,j}$ is not efficient as exponentially many
repetitions of the procedure for $V^{d,+1}_{g,1}$ would be
required.

For many physical systems, efficient implementation of the
$V^{d,+1}_{g,j}$ can be carried out by shifting the procedure for
implementation of $V^{d,+1}_{g,1}$ along path $g$ in $A$ until a
component system in the state $|g(j)\rangle$ is encountered. At
this point implementation of $V^{d,+1}_{g,1}$ is started.

Efficient implementability for the basic arithmetic operations
also implies that there exist Hamiltonians $H^{+}_{g,d}$ and
$H^{\times}_{g,d}$ that efficiently carry out $+_{g,d}$ and
$\times_{g,d}$. Since the definitions of $+$ and $\times$ are
given in terms of the $V^{+1}_{j}$,(Eq. \ref{plus} and the
Appendix), it follows that if the $V^{d,+1}_{g,j}$ can be
efficiently implemented, so can $+_{g,d}$ and $\times_{g,d}$. For
microscopic systems the fact that the times $t_{+},t_{\times}$
required for these implementations are greater than those for the
$V^{d,+1}_{g,j}$ means that the values of $L$ for which
$t_{+}<t_{dec}$ and $t_{\times}<t_{dec}$ may be less than those
possible for just the $V^{d,+1}_{g,j}$.

Another aspect of the efficient implementability condition is
that the thermodynamic resources required to implement
$V^{d,+1}_{g,j}$ must be polynomial and not exponential in $j$.
This takes account of the fact that all computations occur in a
noisy environment and one must spend thermodynamic resources to
protect the system from errors. This is especially the case for
quantum computation for which entanglements of states that
develop as the computation progresses must be protected from
decoherence \cite{Zurek,Unruh,Brandt}. Methods of protecting
these states include the use of quantum error correction codes
\cite{QCerrors} and possibly generation and use of EPR pairs
\cite{Bennett}. These considerations are another reason why it is
important to minimize the time required to implement
$V^{d,+1}_{g,j}$.

There are many physical systems where  the resources needed to
implement $V^{d,+1}_{g,j}$ (other than those involved in the
shift) are either independent of $j$ or are at most polynomial in
$L$. The needed resources do not depend exponentially on $j$ or
$L$. These systems satisfy the requirement of efficient
implementability. There are others that do not. Consider, for
example, a 1-D lattice of systems where the intensity of
environmental interference and noise grows exponentially with
$j$. Here the thermodynamic resources needed to protect the
system from decoherence, etc., would grow exponentially with
$j$.  Another simpler type of system that would be excluded would
be a row of isolated harmonic oscillator potentials each
containing a single spinless particle. The proposed two qubit
states are the ground and first excited states in the well.
However the spring constants of the wells depend exponentially on
$j$. For example the spring constant $p(j+1)$ of the $j+1st$ well
is related to that for the $jth$ well by $p(j+1)=kp(j)$.

For networks of quantum gates efficient implementation of the
basic arithmetic operations has two components. The number of
quantum gates (or steps) in the network must be polynomial in
$L$, as in \cite{Beckman,Vedral}, and the resources needed to
implement individual quantum gates must be polynomial in the
locations of the individual systems addressed by each gate. In
the physical models described above, this second requirement is
not satisfied as resources needed to implement a quantum gate
between the $jth$ and $j^{\prime}th$ qubits depend exponentially
on $j$ and $j^{\prime}$. The fact that one would not build such
models or could not build such models for large $L$ is not
relevant here.

The condition of efficient implementability also places
restrictions on the values of $k$ allowed for $k-ary$
representations. In general values of $k$ are used that are quite
small (e.g. $k=2,\; k=10$, etc.). Except for special cases, $k=1$
(unary) representations are excluded as arithmetic operations are
exponentially hard. Also the value of $k$ cannot be too large.
One reason is that there are physical limitations on the amount
of information that can be reliably stored and distinguished per
unit space time volume \cite{Lloyd}. Also the requirement of
efficient implementation enters in that for large $k$ (e.g.
$k=10^{6}$), even a simple process such as adding two single
digit numbers becomes quite lengthy.

\section{Is the Model ${\cal H}^{arith}$ Necessary?}
\label{IMHN} The preceding was based on first constructing a
purely mathematical Hilbert space model ${\cal H}^{arith}$for
mdoular arithmetic and then using this to construct a physical
model on a space ${\cal H}^{phy}$ that has the same tensor
product structure as ${\cal H}^{arith}$.  The question arises if
the purely mathematical model based on ${\cal H}^{arith}$ is
necessary.  Can one go directly from the axioms of modular
arithmetic to physical models without the use of the model based
on ${\cal H}^{arith}$?

In general this is possible as any structure, physical or
mathematical, that satisfies the axioms is acceptable.  However,
the intermediate mathematical models serve as a useful reference
point for discussions.  This is clear from the literature in
which much use is made of such a model.  For instance any
reference to product qubit states $|\underline{0}\rangle
,|0110110\cdots\rangle$, etc. and linear superpositions of these
states is implicitly using a model based on ${\cal H}^{arith}$.

Another point, already noted, is that the axioms of arithmetic,
modular or not, make no mention of efficient implementability.
Models based on unary representations are just as valid as are
any others.  This is true even if additional axioms are added
giving the properties of the $V^{+1}_{j}$ operators.

This raises the following questions:  Suppose one starts with an
arbitrary quantum system with states in a space ${\cal H}^{phy}$
whose tensor product structure (if any) is unknown. Can
operators, indexed by values in a set of physical parameters for
the system, be defined with properties such that they satisfy the
axioms of modular arithmetic? As will be seen in the following,
this seems possible.  If one also requires that the operators and
those for the basic arithmetic operations be efficiently
implementable, does it follow that ${\cal H}^{phy}$ must have a
tensor product structure based on the defined operators and their
properties?  At present, the answer is not known.

To be specific, the interest is in constructing a model of
arithmetic $\bmod k^{L}$ directly on the state space ${\cal
H}^{phy}$ of a quantum system where ${\cal H}^{phy}$ has an
arbitrary tensor product structure. A set $A$ of $L$ operators
$V_{a}$ on ${\cal H}^{phy}$ indexed by the physical parameters
$a\epsilon A$ is required to have properties that are necessary
conditions for ${\cal H}^{phy}$ to have the tensor product
structure suitable for length $L$ $k-ary$ representations  of
numbers.  These properties are,

\begin{enumerate}
\item Each $V_{a}$ is a cyclic shift.
\item The $V_{a}$ all commute with one another.
\item For each $a \epsilon A$, if $(V_{a})^{k} \neq 1$ there is a unique $a^{\prime}\neq a$ such
that $(V_{a})^{k} = V_{a^{\prime}}$.
\item For each $a^{\prime}$ , if there is an $a\neq a^{\prime}$ such that
$(V_{a})^{k} = V_{a^{\prime}}$, then $a$ is unique.
\item There is just one $a$ for which $(V_{a})^{k}=1$.
\item For just one $a$ there are no $a^{\prime}$ such that
$(V_{a^{\prime}})^{k}=V_{a}$.
\end{enumerate}

The properties reflect those possessed by the $V^{+1}_{j}$, note
especially Eq. \ref{logeff}. Properties 3-6 can be used to
establish a numbering of the label set $A$ with the maximum and
minimum labels given by properties 5 and 6. The commutativity and
cyclic shift properties \cite{Halmos} give the existence of of a
set $\cal B$ of pairwise orthogonal subspaces of states such that
for each $a$ and each subspace $\beta$ in $\cal B$, $V_{a}\beta$
is in $\cal B$ and is orthogonal to $\beta$. In the special case
that the subspaces in $\cal B$ are one dimensional, the subspaces
$\beta$ in $\cal B$ correspond to pairwise orthogonal states
$|\beta\rangle$ such that for each $|\beta\rangle$ in $\cal B$,
$V_{a}|\beta\rangle$ and $|\beta\rangle$ are orthogonal.

One can use property 3 along with iterations $(V_{a})^{h}$ for
$h=0,1,\cdots k-1$ for each $a$ to generate a cyclic ordering or
numbering of the states in $\cal B$ and show that the set
contains $k^{L}$ states. However none of this is sufficient to
select a state as the zero state.  This must be done by making an
arbitrary choice.
\begin{description}
\item[     7.] There is a unique state $|\beta_{\underline{0}}\rangle$ in $\cal B$ which is
the zero state.
\end{description}

Based on this choice one can associate with each string of
numbers, $n_{L},n_{L-1},\cdots n_{\ell},\cdots
,n_{2},n_{1}=\underline{n}$ with $0\leq n_{\ell}\leq k-1$ for each
$\ell$ a unique state $|\beta_{\underline{n}}\rangle$. The
association is given by
\begin{displaymath}
|\beta_{\underline{n}}\rangle =
\prod_{\ell=1}^{L}(V_{a_{\ell}})^{n_{\ell}}
|\beta_{\underline{0}}\rangle.
\end{displaymath}
where the properties of the $V_{a}$ show that the states
$|\beta_{\underline{n}}\rangle$ for different number strings
$\underline{n}$ are orthogonal.

The above can also be used to define addition as in Eq.
\ref{plus} and show that $|\beta_{\underline{0}}\rangle$ is the
additive identity. This and use of the discussion in Section
\ref{AHSM} suggests that these operators and the associated
states do satisfy the axioms of  arithmetic $\bmod k^{L}$. However
examples can be constructed to show that it is very unlikely that
the existence of operators with these properties are sufficient
conditions for ${\cal H}^{phy}$ to have a  tensor product
structure suitable for $k-ary$ representations of length $L$. If
one adds the additional requirement that these operators be
efficiently implementable, then it is an open question if all
these conditions are sufficient to require that ${\cal H}^{phy}$
has a tensor product structure suitable for $k-ary$
representations of length $L$.

\section{Discussion}
\label{Disc} Several points about the work done here should be
noted. The state descriptions of composite quantum systems used in
this paper have not taken account of whether or not the component
systems are distinguishable by properties other than those
explicitly shown in the states.  This is based on the
consideration that the only properties used by a quantum
algorithm are those expressed explicitly in the states and
operators representing the basic arithmetic operations. For
indistinguishable systems, it is suspected that taking account of
their bosonic or fermionic nature, as has been done elsewhere
\cite{BrKi,Vlasov}, will not change the results obtained.
However, this must be investigated.

The condition of efficient implementation of the basic arithmetic
operations is the main restrictive condition on states of quantum
systems that represent numbers. As noted it excludes $k=1$ and
large $k$. It also greatly restricts which unitary operators from
${\cal H}^{arith}$ to ${\cal H}^{phy}$ are allowed. To see this
note that any unitary operator $U$, tensor product preserving or
not, from ${\cal H}^{arith}$ to ${\cal H}^{phy}$ gives a model of
the axioms of modular arithmetic on ${\cal H}^{phy}$. The numbers
are represented by the states $U|\underline{s}\rangle$ and the
basic operators by $UV^{+1}_{j}U^{\dagger}$ and $(U\otimes
U)+(U^{\dagger}\otimes U^{\dagger})$ and similarly for $\times$.
However most of these $U$ can be excluded because the
corresponding basic operators on ${\cal H}^{phy}$ are  not
efficiently implementable. Also for most $U$ there is no way to
physically prepare the states $U|\underline{s}\rangle$. This is
the main reason for the restriction that $U$ be tensor product
preserving with the form of $W_{g,d}$.

Unfortunately there is no way to define exactly which $U$
operators are allowed and which are not.  The reason is that there
is no way to precisely define the meaning of physical
realizability. One needs an hypothesis  for physical
realizability equivalent to the Church-Turing Hypothesis
\cite{Church,Nielsen,Deutsch} for computable functions. Earlier
attempts to characterize realizable physical procedures as
collections of instructions \cite{FouRan,Eks}, or state
preparation and observation proceedures as instruction booklets
or programs for robots \cite{BenEks} have not been generally
accepted.  This problem also arises in describing exactly the
class of tasks that a quantum robot \cite{BenQR} can carry out.

Another aspect of the representation of numbers by quantum states
is that the sets of numbers $1,\cdots ,L$ and $0,\cdots ,k-1$ have
been used to describe $k-ary$ representations of numbers of
length $L$ by quantum states.  For example numbers in either of
these sets are used to describe the $V^{+1}_{j}$ operations. Also
the definitions of $+$ and $\times$ were given in terms of
numbers of iterations  of $V^{+1}_{j}$ and $+$ respectively.

Two components of this should be noted.  One is that the role of
these numbers is limited to the dynamical implementation of the
$V^{d,+1}_{g,j},\; +_{g,d}$, and $\times_{g,d}$. For example, any
method based on a Hamiltonian $H_{g}^{d}$ that implements
$V^{d,+1}_{g,j}$ as a translation of a procedure for implementing
$V^{d,+1}_{g,1}$ by $j$ sites along $g$ requires  motion along
$g$ until the site $g(j)$ is reached. This can be done by
repeated subtraction of $1$ from $j$, interleaved with motion of
some system, such as a head or quantum robot \cite{BenQR}, along
$g$ until $g(j)$ is reached. Also the "carry $1$" operation,
which is part of $V^{d,+1}_{g,j}$ means that motion along the
remaining $L-j$ elements of path $g$ must be built into
$H_{g}^{d}$.

Similar arguments apply for the efficient carrying out of the
$+_{g,d}$ operation as this requires up to $k$ iterations of
$V^{d,+1}_{g,j}$ for each $j$. One method of implementation
requires interleaving the implementation of a procedure for
$V^{+1}_{g,j}$ with subtractions of $1$ from a state
$|\underline{s}_{j}\rangle$, Eq. \ref{plus}, until
$|\underline{0}_{j}\rangle$ is obtained.

Implementation of these operations by quantum systems means that
numbers up to $L$ and $k$ must also be represented by quantum
states of systems.  These systems can either be mobile and part
of the head or fixed external systems. Thus the  arguments and
conditions  already discussed apply to these representations too.

The other component is that the magnitudes of the numbers
represented by the states of systems that are part of the
dynamics  are exponentially smaller than those represented by the
system  on which the dynamics is acting.  States of a composite
quantum system satisfying the conditions for  $k-ary$ number
representations of length $L$, represent the first $k^{L}$
numbers. Numbers appearing in the dynamics range up to $k$ and
$L=\log_{k}k^{L}$. This exponential decrease is a consequence of
the requirement of efficient implementability of arithmetic
operations.

The conditions discussed in this paper, including the requirement
of efficient physical implementability, also apply to the quantum
states of ancillary systems that are used to implement the
dynamics of an algorithm.  This is  evident in any algorithm
which interleaves evaluation of some numerical function with
carrying out an action until a specified function value is
reached. For instance, implementation of the $V^{d,+1}_{g,j}$,
e.g. by use of a head or quantum robot with an on board quantum
computer \cite{BenQR}, would require a quantum computer with at
least $O([\log_{m}{(L)}]+1)$ qubytes for an $m-ary$
representation of numbers up to $L$. ($[-]$ denotes the largest
integer in.) Here the dynamics that carries out these operations
is subject to all the requirements described so far. It is also
part of the dynamics for implementing $V^{d,+1}_{g,j}$.

These considerations suggest that it may not be possible to
describe the representation of numbers by states of a composite
quantum system without the use of states of other systems already
assumed to represent numbers.  These states are part of the
dynamics of the basic arithmetic operations.

Whether this is true or not is a question for the future.
However, if this impossibility is the case, one is helped by the
fact that the number of states needed to represent numbers in the
dynamics is exponentially smaller than the number of states
representing numbers of the composite system on which the dynamics
acts.

Finally it should be noted that much of the discussion, including
the efficient implementability condition, which has been applied
to microscopic quantum systems, also applies to macroscopic
quantum systems.  In this case $t_{dec}\ll t_{sw}$ so the
limitation that the number of steps is $<t_{dec}/t_{sw}$ is not
applicable. Instead efficient implementation means that there
exists a dynamics such that the number of steps needed to carry
out arithmetic operations is polynomial in $L$.  Also the states
of the system used to represent numbers are those that are
stabilized by the interactions with the environment, the "pointer
states" \cite{Zurek1,Zeh,Venugopalan}. The fact that these
conditions are much less onerous than the limitations on
microscopic systems is shown by the widespread use of macroscopic
computers and counting devices and timers.

In conclusion it is reemphasized that this work is one approach
to making explicit the assumptions and conditions involved in the
representation of natural numbers by states of quantum systems. It
is based on separating the mathematical concept of numbers, as
models of a set of axioms, from the physical concept of efficient
implementabiliy of the basic arithmetic operations described by
the axioms. Whether this approach will turn out to be a good one
or not depends on future work.

\section*{Acknowledgements}
Discussions with Murray Peshkin on several points of this paper were much appreciated.
This work is supported by the U.S. Department of Energy, Nuclear Physics Division,
under contract W-31-109-ENG-38.

\section*{Appendix: Definition of $\times$}
The goal is to define a  unitary times operator according to Eq.
\ref{times} based on efficient iteration of the $+$ operator. To
this end define $Q_{j}(2,3)$ for $j=1,\cdots ,L$ as operators on
the second and third product states that convert
$|\underline{s},\underline{w},\underline{w0^{j-1}},\underline{z}
\rangle$ to $|\underline{s},\underline{w},\underline{w0^{j}},
\underline{z}\rangle$. It has the effect of multiplying
$|\underline{w0^{j}}\rangle$ by $k$.  An efficient reversible
implementation of this, acting on the state
$|\underline{s},\underline{w},\underline{y},\underline{z}\rangle$
is obtained by subtraction, $\bmod k$, of the $L-j+1st$ component
qubyte state of $|\underline{w}\rangle$ from the $Lth$ component
state of $|\underline{y}\rangle$, shifting all the elements of
$|\underline{y}\rangle$ by one site and putting the result of the
subtraction at the newly opened first site. This works because, if
$|\underline{y}\rangle=|\underline{w0^{j-1}}\rangle$, then
$|\underline{y}_{L}=|\underline{w}_{L-j+1}\rangle$. The result,
$|\underline{0}_{L}\rangle$, of the subtraction is moved to the
first site of $|\underline{y}\rangle$ after the shift. One has
\begin{equation}
Q_{j}(2,3)
|\underline{s},\underline{w},\underline{y},\underline{z} \rangle =
|\underline{s},\underline{w},\underline{y^{\prime}},
\underline{z}\rangle
\end{equation}
where $|\underline{y^{\prime}_{j+1}}\rangle
=|\underline{y}_{j}\rangle$ for $1\leq j \leq L-1$ and
$|\underline{y^{\prime}_{1}}\rangle = |\underline{y}_{L}\rangle
\ominus |\underline{w}_{L-j+1}\rangle$.  Here $\ominus$ denotes
subtraction $\bmod k$.  Note that $Q_{j}(2,3)$ is unitary.

The operator $\times$ is defined from the $Q_{j}(2,3)$ and $+$ by
\begin{eqnarray*}
 & \times |\underline{s},\underline{w},\underline{y},
\underline{z}\rangle = Q_{L}(2,3)(+_{3,4})^{s_{L}} Q_{L-1}(2,3)
(+_{3,4})^{s_{L-1}} \nonumber \\
&  \cdots ,(+_{3,4})^{s_{2}} Q_{1}(2,3)(+_{3,4})^{s_{1}}+_{2,3}
|\underline{s},\underline{w},\underline{y},\underline{z}\rangle
\end{eqnarray*}
Here $+_{m,n}$ carries out the action defined in Eq. \ref{plus}
on the $mth$ and $nth$ product state. The $mth$ state remains
unchanged in this action. $s_{h}$ is the number $
\underline{s}(h)$ in the state component
$|\underline{s}(h),h\rangle$ of $|\underline{s}\rangle$.  Note
that since each operator in the righthand product of the equation
is unitary, so is $\times$.

To see that $\times$ as defined above does carry out the intended
multiplication operation on initial states of the form
$|\underline{s},\underline{w},\underline{0},
\underline{0}\rangle$ one carries out the action of the $2L+1$
operators shown above. The steps give
\begin{eqnarray}
& |\underline{s},\underline{w},\underline{0},\underline{0}\rangle
\begin{array}{l} +_{2,3}  \\ \longrightarrow \end{array}
|\underline{s},\underline{w},\underline{w}, \underline{0}\rangle
\begin{array}{c} (+_{3,4})^{s_{1}} \\ \longrightarrow \end{array}
|\underline{s},\underline{w},\underline{w}, \underline{s_{1}w}
\rangle \nonumber \\ & \begin{array}{c} Q_{1}(2,3) \\
\longrightarrow
\end{array} |\underline{s},\underline{w},\underline{w0},
\underline{s_{1}w} \rangle  \begin{array}{c} (+_{3,4})^{s_{2}}
\\ \longrightarrow \end{array}
|\underline{s},\underline{w},\underline{w0},
\underline{s_{1}t+s_{2}t0} \rangle \nonumber \\
& \cdots\begin{array}{c} Q_{L}(2,3) \\ \longrightarrow
\end{array}
|\underline{s},\underline{w},\underline{0},
\underline{s_{1}w+s_{2}t0+ \cdots + s_{L}t0^{L-1}} \rangle
\nonumber
\end{eqnarray}
Note that $Q_{L}(2,3)$ acting on $|-
,\underline{w},\underline{w0^{L-1}}, -\rangle$ gives $|-
,\underline{w},\underline{0},-\rangle$ in accordance with Eq.
\ref{times} as $|\underline{w0^{L}}\rangle =
|\underline{0}\rangle$.  Here $|\underline{s_{1}w}\rangle$
denotes $s_{1}$ iterations of adding $|\underline{w}\rangle$ to
$|\underline{0}\rangle$; also $s_{j}w0^{j-1}$ denotes the result
of $s_{j}$ additions of $|\underline{w0^{j-1}}\rangle$ to the
$4th$ product state.

\end{document}